\documentstyle[twocolumn,prl,aps,epsf]{revtex}
\begin{document}
\twocolumn[\hsize\textwidth\columnwidth\hsize\csname
@twocolumnfalse\endcsname
\title{Co-Evolution of quasispecies:\\ 
B-cell mutation rates maximize viral error catastrophes}
\author{Christel Kamp and Stefan Bornholdt\cite{email}}
\address{Institut f\"ur Theoretische Physik, Universit\"at Kiel,
Leibnizstra{\ss}e 15, D-24098 Kiel, Germany}
\maketitle
\date{today}
\begin{abstract}
Co-evolution of two coupled quasispecies is studied, motivated by 
the competition between viral evolution and adapting immune response. 
In this co-adaptive model, besides the classical error catastrophe 
for high virus mutation rates, a second ``adaptation-'' catastrophe
occurs, when virus mutation rates are too small to escape immune attack. 
Maximizing both regimes of viral error catastrophes is a possible 
strategy for an optimal immune response, reducing the range of allowed 
viral mutation rates to a minimum. From this requirement, one obtains 
constraints on B-cell mutation rates and receptor lengths, yielding 
an estimate of somatic hypermutation rates in the germinal center 
in accordance with observation.   
\medskip \\ 
\end{abstract}
]

During the past 30 years, the concept of quasispecies 
\cite{eigen71,eigen79} 
has developed into a valuable tool for modeling key features of 
molecular and viral evolution. It draws a simple picture of 
sequence evolution in the presence of a peaked fitness function,
resulting in the formation of a central master 
sequence surrounded by a cloud of mutant sequences, summarized 
as the quasispecies. 
A prominent feature of such systems is the occurrence of an 
error catastrophe, a sudden breakdown of stability when mutation 
rates get large. Recent developments of quasispecies models  
include their formulation within a statistical mechanics 
framework and a characterization of the error catastrophe 
as a phase transition
\cite{nowak89,taraz92,bonh93,baake97,altm01}. 

While traditionally defined on static fitness landscapes, the 
concept of quasispecies recently has been extended to non-stationary 
fitness environments \cite{nilsson00}. This has important 
applications, as viruses usually face quickly changing environments 
in the tight and temporal niches of their hosts. This new approach 
allows for studies of the adaptive response of a quasispecies to 
changing external conditions. This has been studied for different 
choices of time dependent fitness functions \cite{wilke99,nilsson00}. 

Here, we further extend this approach and study  
a virus in the environment of an adaptive immune system. 
To include the adaptive response of an immune system
into the model, a further step has to be taken beyond
a simple time dependent viral fitness function. 
An interesting observation is that motifs of immune 
receptors often form something very similar to quasispecies: 
A ``master'' sequence of the immune receptor, corresponding 
to the complementary epitope motif of the virus, is surrounded 
by a cloud of closely related receptor sequences that emerge 
from somatic hypermutation in the germinal centers 
\cite{harris99,meyer01}. 
As the presence of the viral motif induces the proliferation 
of the corresponding receptor sequence, one can argue that 
this process can be represented in the framework of quasispecies,
interpreting the conditions that lead to proliferation of 
specific immune receptors as the fitness of the particular 
receptors. Let us, therefore, formally consider the B-cell 
population induced by a specific viral epitope as a 
quasispecies itself. 
In this paper we, therefore, study the co-evolution of two 
asymmetrically coupled quasispecies under competition. 
While the immune quasispecies is strongly attracted by the virus, 
the viral quasispecies is driven away from its current master
sequence by the immune system. This results in a migration 
through sequence space as observed in 
many infectious diseases as in HIV \cite{ganeshan97,allen00}.
In the following, let us first define the model in detail. 
Then, dynamical regimes and stability bounds are 
discussed which occur as a result of the selective forces 
acting on both sides of the system. Finally, from the perspective 
of an optimal immune response, a relationship between receptor 
size and mutability is derived. By maximizing the ranges of 
error catastrophes for the virus, bounds on observable 
mutation rates in the immune system are derived. 

Consider a model with two quasispecies of genetic sequences,  
one of them coding for a virus and the second one coding for 
the variable part of an immune receptor. Sequence lengths are 
$n_v$ and $n_{is}$, respectively, with bases taken from an 
alphabet of length $\lambda=4$. Mutation rates are quantified 
by the copy fidelities per base $q_v,q_{is}<1$. The time 
evolution of the two distributions, the concentration $z_k(t)$
of viral sequence $k$, as well as the concentration of immune 
cells $y_k(t)$, representing receptor coding sequence $k$,   
is described by two sets of coupled differential equations
of the type introduced by Eigen \cite{eigen71}
\begin{eqnarray}
\dot{y}_k &=& \sum_l \; W^{is}_{kl} \; A(z_l) \; y_l 
           - f \; y_k \label{eigen1} \\
\dot{z}_k &=& \sum_l \; (W^v_{kl} \; B_l 
           - \delta_{kl} \; C(y_l)) \; z_l \label{eigen2} \\
W^i_{kl}  &=& \frac{q_i^{n-d(k,l)}(1-q_i)^{d(k,l)}}{(\lambda-1)^{d(k,l)}}
\smallskip \\ 
&& i \in \{is,v\}; \ \ \ \ k,l\in\{1,...,\lambda^n\}.
\nonumber
\end{eqnarray}
Locally in sequence space, we assume a simple one-to-one map 
between viral sequence and the sequence coding for the immune 
receptor that maximally fits the viral epitope (within the 
local neighborhood of mutated receptors). For simplicity, 
both sequences share the 
same subscript $l$ in the above formulation. Therefore, $A(z_l)$ 
denotes the growth rate of the B-cell clone corresponding to 
receptor sequence $l$ and depends on the concentration of its 
complementary viral sequence $z_l$. The viral replication rate  
is $B_l$, and its decay rate $C(y_l)$, which depends on the 
associated immune cell concentration. 
For viral, as well as immune receptor evolution, a transition 
probability $W^i_{kl}$ from a sequence $l$ to sequence $k$ by 
mutation is assumed, depending on the respective mutation rate 
$q_i$, sequence length $n_i$ and the Hamming distance of the 
two sequences $d(k,l)$. For simplicity let us assume 
$n_{is}=n_v=n$ with the comparable complexity of viral 
epitopes and corresponding immune receptors in mind. 
Equation (\ref{eigen1}) models the relative concentrations
of immune receptor coding sequences, with constant overall 
population size normalized by $f=\sum_l \; A(z_l) \; y_l$. 
The viral population (\ref{eigen2}), on the other hand, 
does not reach a constant population size, 
as the immune system usually works efficiently enough 
for the virus not to enter the regime of saturation. 
Therefore, absolute concentration is considered here 
and is the adequate quantity to quantify viral feed-back 
to immune cell proliferation.

As viral existence in sequence space often is restricted to narrow 
niches with high fitness, let us assume a fitness landscape
with a single peak with $B_l=\sigma_v\gg B_{m\neq l} = \eta_v $ 
that moves as a result of immune system pressure. Vice versa, 
as the immune response represents a very specific answer to a pathogen,
let us also model the immune fitness landscape to have a single peak, 
corresponding to the receptor matching the current antigenic 
master sequence. This simplifies the co-evolution model  
to a few discrete alternatives: 
$A(z_l)=\sigma_{is}$ if and only if $z_l$ represents
the concentration of the viral master sequence, otherwise 
$A(z_l)=\eta_{is}\ll\sigma_{is}$. Analogously, $C(y_l)=\delta$
if and only if $y_l$ represents the dominant immune
receptor's concentration, otherwise $C(y_l)=0$. 
This makes the complicated couplings in the above 
equations tractable and allows us to neglect mutational 
backflow to the respective master sequences. 
Let us first write down simplified equations that apply 
to both quasispecies (1) and (2). For this purpose we 
use non-normalized concentrations \cite{jones76,thompson74}
for quasispecies (1) also, and neglect the decay term in (2)
for the moment. Then, each of the two quasispecies can be 
written in terms of the concentrations of a master sequence 
$x_0$ and of an arbitrary sequence of the first error 
class $\Gamma_1$ $x_1$.
\begin{eqnarray}
\dot{x}_0(t)&=&q^n\sigma x_0(t)\label{dgl1}\\
\dot{x}_1(t)&=&\frac{1-q}{\lambda-1}q^{n-1}\sigma x_0(t)
+ q^n\eta x_1(t)\label{dgl2}
\end{eqnarray}
The interaction between the two systems has to be specified 
by extra rules on the basis of the above definition of growth
and decay rates. To keep the model as simple as possible, 
the decay rate $\delta$ only affects the viral sequence 
matching the dominant immune receptor. If this happens to 
coincide with the viral master sequence, the viral fitness 
peak will effectively move. Depending on its strength, 
the former fitness peak eventually will drop below the 
environmental growth rate. To be specific, the dynamical 
rules of this process are defined as follows:

\begin{enumerate}
\item
Once the immune system imposes a decay rate $\delta>0$ on 
the viral master sequence (so far stabilized at the viral 
fitness peak), the narrow niche of the virus is assumed to 
move to an arbitrary sequence of the first error class. 
\item
The viral quasispecies adapts to the new fitness peak on 
a time scale $\tau_v$ given by the dynamical equations above. 
\item
The fitness peak of the immune quasispecies is adjusted 
to the new maximum of the viral distribution. 
\item
The immune system adapts to the new fitness peak on a 
second time scale $\tau_{is}$ determined as above. 
\end{enumerate}

These steps are then iterated. While this is a strongly 
simplified picture of the co-evolutionary dynamics of two
coupled quasispecies, the localization of the interaction 
to their respective master sequences, as well as the restriction 
to only two growth rates for each quasispecies, allow a simple 
estimate of the dynamical regimes of the two coupled sets of 
equations of type (\ref{dgl1},\ref{dgl2}). Each of 
the fitness peaks is adjusted once during each cycle 
of duration $\tau=\tau_v+\tau_{is}$ (in steps 1 resp.\ 3). 
This allows us to follow the arguments of Nilsson and Snoad 
\cite{nilsson00} to determine the relative growth of the 
respective future master sequences over a full cycle $\tau$ 
as a criterion for the quasispecies' survival \cite{nilsson00}
\begin{equation}\label{kappa}
\kappa
=\frac{1}{e^{\eta\tau}}\frac{x_1(\tau)}{x_0(0)}
=\frac{\left(e^{(q^n\sigma-\eta)\tau}-e^{(q^n\eta-\eta)\tau}\right)
(1-q)\sigma}{(\lambda-1)(\sigma-\eta)q}.
\end{equation}
This expression is applied to each one of the two quasispecies 
(with the respective variables), defined over the full interval 
$\tau=\tau_v+\tau_{is}$ between two adjustments of its fitness peak. 
For a relative growth coefficient $\kappa>1$ a species will survive 
and for $\kappa\leq 1$ it will get extinct.

Now consider a coupled system of viral and immune quasispecies
where the immune part exerts a selective pressure on the virus in the 
form of a non-vanishing kill or decay rate $\delta$. To estimate the 
migration time scale of the virus $\tau_v$, let us iterate the model 
for a full cycle $\tau$ starting at the moment of the move of the 
fitness peak at $t=0$. The relative size of the old and new master 
sequence peaks is then subsequently determined for another time 
interval $\tau_v$. Let us assume $x_1(0)=0$ since the new error class 
one sequence members are mainly recruited from the former, weakly 
populated error class two. The time scale $\tau_v$ is given by the 
waiting time until the new master sequence population exceeds the old one:  
\begin{eqnarray}
e^{(q_v^n\eta_v-\delta)\tau_v}x_0(\tau)&\stackrel{!}{=}
&e^{q_v^n\sigma_v\tau_v}x_1(\tau)\\
\Rightarrow e^{(q_v^n\eta_v-\delta)\tau_v}e^{q_v^n\sigma_v\tau}
\hspace{-0.1cm}&=&
e^{q_v^n\sigma_v\tau_v}\frac{\left(e^{q_v^n\sigma_v\tau}\hspace{-0.12cm}
- e^{q_v^n\eta_v\tau}\right)\hspace{-0.1cm}(1\hspace{-0.08cm}
-q_v)\sigma_v}{(\lambda-1)(\sigma_v-\eta_v)q_v}.
\nonumber 
\end{eqnarray}
Mutational flows between the involved sequences can be neglected 
due to the small growth of the former master sequence and the small 
size of the initial new master sequence population. 
Assuming $\sigma_v\gg\eta_v$ and $q_v \approx 1$ 
the viral adaptation timescale can be estimated to   
\begin{equation}\label{tv}
\tau_v\approx - \frac{\ln\left(\frac{1-q_v}{\lambda-1}\right)}
{q_v^n(\sigma_v-\eta_v)+\delta}. 
\end{equation}
Similarly, for the migration time for the immune quasispecies 
$\tau_{is}$ we obtain 
\begin{eqnarray}
\tau_{is} \approx -\frac{\ln\left(
\frac{1-q_{is}}{\lambda-1}\right)}{q_{is}^n
(\sigma_{is}-\eta_{is})}.
\end{eqnarray}
Both, $\tau_v$ as well as $\tau_{is}$, exhibit a local minimum 
at specific values of their copy fidelities $q_v$ and $q_{is}$, 
mainly determined by the balance between the requirement of 
a sufficiently large initial population for the formation of 
a future master sequence and sufficiently low mutational
losses of the new master sequence. Inserting $\tau$ into the 
expressions for viral stability $\kappa_v$ and immune stability 
$\kappa_{is}$ according to (\ref{kappa}), one obtains estimates 
for the regimes of viral and immune (co-)existence.

\begin{figure}[hbt]
\vspace{-1.5cm}
\setlength{\unitlength}{0.240900pt}
\ifx\plotpoint\undefined\newsavebox{\plotpoint}\fi
\begin{picture}(1200,1170)(130,0)
\font\gnuplot=cmr10 at 10pt
\gnuplot
\sbox{\plotpoint}{\rule[-0.200pt]{0.400pt}{0.400pt}}%
\put(414,547){\makebox(0,0)[l]{{\small Virus $-$ }}}
\put(414,433){\makebox(0,0)[l]{{\small Virus $+$ }}}
\put(785,776){\makebox(0,0)[l]{{\small IS  $-$}}}
\put(785,662){\makebox(0,0)[l]{{\small IS $+$}}}
\put(384,547){\vector(-1,0){62}}
\put(587,547){\vector(1,0){62}}
\put(671,382){\usebox{\plotpoint}}
\put(671.00,382.00){\usebox{\plotpoint}}
\put(655.92,396.20){\usebox{\plotpoint}}
\put(640.73,410.27){\usebox{\plotpoint}}
\put(624.89,423.67){\usebox{\plotpoint}}
\put(608.99,437.01){\usebox{\plotpoint}}
\put(597.35,452.10){\usebox{\plotpoint}}
\multiput(596,467)(-1.479,20.703){2}{\usebox{\plotpoint}}
\put(595.05,513.78){\usebox{\plotpoint}}
\put(604.00,532.50){\usebox{\plotpoint}}
\put(611.87,551.70){\usebox{\plotpoint}}
\put(620.05,570.78){\usebox{\plotpoint}}
\put(631.29,587.94){\usebox{\plotpoint}}
\put(642.60,605.34){\usebox{\plotpoint}}
\put(653.43,623.04){\usebox{\plotpoint}}
\put(665.12,640.17){\usebox{\plotpoint}}
\put(677.63,656.72){\usebox{\plotpoint}}
\put(690.65,672.87){\usebox{\plotpoint}}
\put(702.63,689.81){\usebox{\plotpoint}}
\put(715.64,705.86){\usebox{\plotpoint}}
\put(728.57,722.09){\usebox{\plotpoint}}
\put(741.64,738.19){\usebox{\plotpoint}}
\put(754.09,754.79){\usebox{\plotpoint}}
\put(767.34,770.72){\usebox{\plotpoint}}
\put(780.54,786.72){\usebox{\plotpoint}}
\put(794.06,802.41){\usebox{\plotpoint}}
\put(806.51,819.01){\usebox{\plotpoint}}
\put(820.02,834.70){\usebox{\plotpoint}}
\put(833.41,850.55){\usebox{\plotpoint}}
\put(846.93,866.24){\usebox{\plotpoint}}
\put(859.50,882.75){\usebox{\plotpoint}}
\put(867,891){\usebox{\plotpoint}}
\put(327.00,382.00){\usebox{\plotpoint}}
\put(332.27,402.04){\usebox{\plotpoint}}
\put(338.60,421.80){\usebox{\plotpoint}}
\put(345.89,441.23){\usebox{\plotpoint}}
\multiput(348,445)(0.324,20.753){3}{\usebox{\plotpoint}}
\multiput(347,523)(-2.743,20.573){3}{\usebox{\plotpoint}}
\put(337.85,585.05){\usebox{\plotpoint}}
\multiput(335,601)(-3.507,20.457){2}{\usebox{\plotpoint}}
\multiput(328,639)(-3.237,20.502){2}{\usebox{\plotpoint}}
\put(320.65,687.34){\usebox{\plotpoint}}
\put(317.87,707.91){\usebox{\plotpoint}}
\multiput(316,721)(-2.397,20.617){2}{\usebox{\plotpoint}}
\put(310.48,769.73){\usebox{\plotpoint}}
\multiput(310,775)(-1.987,20.660){2}{\usebox{\plotpoint}}
\put(304.50,831.71){\usebox{\plotpoint}}
\multiput(303,846)(-0.922,20.735){2}{\usebox{\plotpoint}}
\put(301,891){\usebox{\plotpoint}}
\sbox{\plotpoint}{\rule[-0.400pt]{0.800pt}{0.800pt}}%
\sbox{\plotpoint}{\rule[-0.500pt]{1.000pt}{1.000pt}}%
\put(304.00,827.00){\usebox{\plotpoint}}
\put(297,845){\usebox{\plotpoint}}
\put(908.00,717.00){\usebox{\plotpoint}}
\put(887.35,718.28){\usebox{\plotpoint}}
\put(866.70,719.61){\usebox{\plotpoint}}
\put(846.05,720.99){\usebox{\plotpoint}}
\put(825.38,722.00){\usebox{\plotpoint}}
\put(804.70,723.00){\usebox{\plotpoint}}
\put(784.02,724.00){\usebox{\plotpoint}}
\put(763.43,726.00){\usebox{\plotpoint}}
\put(742.76,727.00){\usebox{\plotpoint}}
\put(722.14,728.81){\usebox{\plotpoint}}
\put(701.47,730.00){\usebox{\plotpoint}}
\put(680.83,731.45){\usebox{\plotpoint}}
\put(660.20,733.00){\usebox{\plotpoint}}
\put(639.60,735.00){\usebox{\plotpoint}}
\put(619.00,737.00){\usebox{\plotpoint}}
\put(598.37,738.66){\usebox{\plotpoint}}
\put(577.80,741.00){\usebox{\plotpoint}}
\put(557.24,743.29){\usebox{\plotpoint}}
\put(536.66,745.56){\usebox{\plotpoint}}
\put(516.08,747.85){\usebox{\plotpoint}}
\put(495.60,751.20){\usebox{\plotpoint}}
\put(475.11,754.48){\usebox{\plotpoint}}
\put(454.61,757.73){\usebox{\plotpoint}}
\put(434.11,760.98){\usebox{\plotpoint}}
\put(413.97,765.58){\usebox{\plotpoint}}
\put(393.98,771.01){\usebox{\plotpoint}}
\put(374.20,777.27){\usebox{\plotpoint}}
\put(354.68,784.28){\usebox{\plotpoint}}
\put(336.41,794.06){\usebox{\plotpoint}}
\put(319.75,806.25){\usebox{\plotpoint}}
\put(306.97,822.54){\usebox{\plotpoint}}
\put(304,827){\usebox{\plotpoint}}
\sbox{\plotpoint}{\rule[-0.200pt]{0.400pt}{0.400pt}}%
\put(908.0,318.0){\rule[-0.200pt]{0.400pt}{138.036pt}}
\put(291.0,318.0){\rule[-0.200pt]{148.635pt}{0.400pt}}
\put(291.0,318.0){\rule[-0.200pt]{0.400pt}{138.036pt}}
\put(291.0,891.0){\rule[-0.200pt]{148.635pt}{0.400pt}}
\put(291,278){\makebox(0,0){\small 0}}
\put(291.0,318.0){\rule[-0.200pt]{0.400pt}{4.818pt}}
\put(291.0,871.0){\rule[-0.200pt]{0.400pt}{4.818pt}}
\put(414,278){\makebox(0,0){\small 0.002}}
\put(414.0,318.0){\rule[-0.200pt]{0.400pt}{4.818pt}}
\put(414.0,871.0){\rule[-0.200pt]{0.400pt}{4.818pt}}
\put(538,278){\makebox(0,0){\small 0.004}}
\put(538.0,318.0){\rule[-0.200pt]{0.400pt}{4.818pt}}
\put(538.0,871.0){\rule[-0.200pt]{0.400pt}{4.818pt}}
\put(661,278){\makebox(0,0){\small 0.006}}
\put(661.0,318.0){\rule[-0.200pt]{0.400pt}{4.818pt}}
\put(661.0,871.0){\rule[-0.200pt]{0.400pt}{4.818pt}}
\put(785,278){\makebox(0,0){\small 0.008}}
\put(785.0,318.0){\rule[-0.200pt]{0.400pt}{4.818pt}}
\put(785.0,871.0){\rule[-0.200pt]{0.400pt}{4.818pt}}
\put(908,278){\makebox(0,0){\small 0.01}}
\put(908.0,318.0){\rule[-0.200pt]{0.400pt}{4.818pt}}
\put(600,175){\makebox(0,0){{ $\mu_v$}}}
\put(908.0,871.0){\rule[-0.200pt]{0.400pt}{4.818pt}}
\put(948,318){\makebox(0,0)[l]{\small 0}}
\put(888.0,318.0){\rule[-0.200pt]{4.818pt}{0.400pt}}
\put(291.0,318.0){\rule[-0.200pt]{4.818pt}{0.400pt}}
\put(948,433){\makebox(0,0)[l]{\small 0.002}}
\put(888.0,433.0){\rule[-0.200pt]{4.818pt}{0.400pt}}
\put(291.0,433.0){\rule[-0.200pt]{4.818pt}{0.400pt}}
\put(948,547){\makebox(0,0)[l]{\small 0.004}}
\put(888.0,547.0){\rule[-0.200pt]{4.818pt}{0.400pt}}
\put(291.0,547.0){\rule[-0.200pt]{4.818pt}{0.400pt}}
\put(948,662){\makebox(0,0)[l]{\small 0.006}}
\put(888.0,662.0){\rule[-0.200pt]{4.818pt}{0.400pt}}
\put(291.0,662.0){\rule[-0.200pt]{4.818pt}{0.400pt}}
\put(948,776){\makebox(0,0)[l]{\small 0.008}}
\put(888.0,776.0){\rule[-0.200pt]{4.818pt}{0.400pt}}
\put(291.0,776.0){\rule[-0.200pt]{4.818pt}{0.400pt}}
\put(948,891){\makebox(0,0)[l]{\small 0.01}}
\put(888.0,891.0){\rule[-0.200pt]{4.818pt}{0.400pt}}
\put(1062,605){\makebox(0,0){{ $\mu_{is}$}}}
\put(291.0,891.0){\rule[-0.200pt]{4.818pt}{0.400pt}}
\end{picture}
\vspace{-1.2cm}
\caption{\label{pic1} Regimes of viral and immune quasispecies 
(co-)existence, with $+/-$ denoting stable/unstable regions of 
the respective quasispecies in dependence on mutation rates $\mu_v$ 
and $\mu_{is}$. Parameters are $\sigma_v=\sigma_{is}=10$, 
$\eta_v=\eta_{is}=1$, $\delta=200$, $n_v=n_{is}=50$, 
and $\lambda=4$. A large value of $\delta$ is chosen to get a 
good qualitative view of the system's behavior.}
\end{figure}
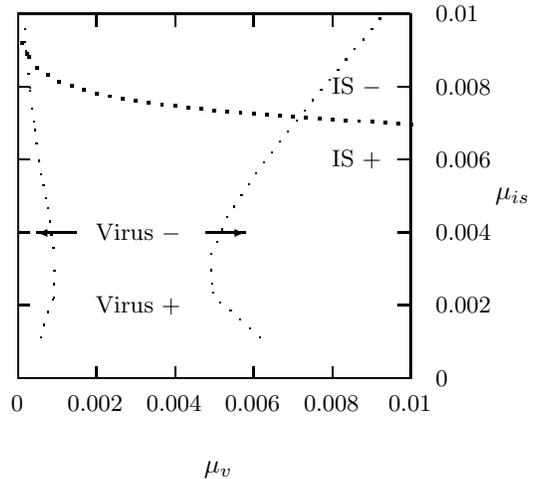

Fig.\ \ref{pic1} shows these regimes in terms of the respective  
mutation rates $\mu=1-q$. The classical error catastrophe occurs 
at lower mutation rates in comparison to the static error threshold 
$\mu^{stat}_{err}=1 - \left(\frac{\eta}{\sigma}\right)^{\frac{1}{n}} 
= 0.045$ (cf. \cite{eigen71}). 
This effect is due to additional mutational losses by migration 
and becomes large for small $\tau$. In addition, Fig.\ \ref{pic1} 
shows different limiting behaviors for $\kappa_v$ and $\kappa_{is}$ 
for $\mu_v\rightarrow  0$ and $\mu_{is}\rightarrow 0$, respectively, 
which can be summarized as 
\begin{eqnarray*}
\kappa_v&\stackrel{\mu_v \rightarrow 0}{\rightarrow}&0 \nonumber \\
\kappa_{is}&\stackrel{\mu_{is}\rightarrow 0}{\approx}
&e^{(\sigma_{is}-\eta_{is})\tau_v}.
\end{eqnarray*}
For the viral quasispecies one observes a second error (``adaptability'') 
catastrophe at small viral mutation rates, because a minimum viral 
mutation rate is needed to escape the decay rate $\delta$ induced 
by the immune response at the viral master sequence. 
The $\kappa_v$-surface in the $\mu_v$-$\mu_{is}$-plane, whose 
$\kappa=0$ contour lines are shown in Fig.\ \ref{pic1}, is dominated 
by a saddle point: $\kappa_v(\mu_v)$ exhibits a local maximum while 
$\kappa_v(\mu_{is})$ shows a local minimum.   
An optimal strategy for viral suppression is, therefore, to 
adjust the mutation rate $\mu_{is}$ of the immune quasispecies 
such that $\kappa_v$ operates in its valley, with maximum regions 
of error catastrophes on both sides. One obtains the condition 
\begin{equation}
\frac{\partial\kappa_v}{\partial \mu_{is}}\stackrel{!}{=}0,
\end{equation}
which can be written as 
\begin{eqnarray}\label{optqis}
\mu_{is}-1+n_{is}\mu_{is}\ln\left(\frac{\mu_{is}}{\lambda-1}\right)&=&0. 
\end{eqnarray}
This mutation rate minimizes the regime of possible existence of the 
viral quasispecies in Fig.\ \ref{pic1}. Depending on the involved 
viral and immune growth rates, this range of allowed viral mutation 
rates $\mu_v$ may vary (and even vanish for some values). 

How does this compare with experimental results? Let us focus on 
B-cells and their antibodies. Each antibody has at least 
two antigen receptors located in the variable regions of the antibody's 
heavy and light chains, each of which contains about 110 amino acids. 
Each receptor is coded by approximately 660 nucleotides. 
Antigen detection takes place in 6 subregions, the complementarity 
determining regions (CDRs) that represent $20-30\%$ of the antibody's 
variable (V-)regions \cite{roitt,aigner}. 
In the course of the primary immune response one observes somatic 
hypermutation in the recombined V-region genes, with mutational 
hot spots at the CDRs, resulting in an enhanced affinity towards 
the invading antigen \cite{harris99,weigert70,nossal92}. 
Observed mutation rates are in the range of $10^{-4}$-$10^{-3}$ 
mutations per base pair per generation \cite{mckean84,berek88,levy89}. 
Mutation rates in the CDRs are approximately twice to tenfold higher 
than those found in the entire V-region \cite{doerner97,berek88}.
These observations are quite universal to adaptive immune systems 
that are common to jawed vertebrates. Differences are mainly found 
in the effectivity of selection due to varying stages of germinal 
centers' expression \cite{diaz98,kasahara98,wilson92,reynaud95,green98}. 
\begin{figure}[hbt]
\begin{center}
\setlength{\unitlength}{0.240900pt}
\ifx\plotpoint\undefined\newsavebox{\plotpoint}\fi
\begin{picture}(1050,720)(60,0)
\font\gnuplot=cmr10 at 10pt
\gnuplot
\sbox{\plotpoint}{\rule[-0.200pt]{0.400pt}{0.400pt}}%
\put(241.0,163.0){\rule[-0.200pt]{4.818pt}{0.400pt}}
\put(221,163){\makebox(0,0)[r]{\small$10^1$}}
\put(1010.0,163.0){\rule[-0.200pt]{4.818pt}{0.400pt}}
\put(241.0,189.0){\rule[-0.200pt]{2.409pt}{0.400pt}}
\put(1020.0,189.0){\rule[-0.200pt]{2.409pt}{0.400pt}}
\put(241.0,223.0){\rule[-0.200pt]{2.409pt}{0.400pt}}
\put(1020.0,223.0){\rule[-0.200pt]{2.409pt}{0.400pt}}
\put(241.0,241.0){\rule[-0.200pt]{2.409pt}{0.400pt}}
\put(1020.0,241.0){\rule[-0.200pt]{2.409pt}{0.400pt}}
\put(241.0,249.0){\rule[-0.200pt]{4.818pt}{0.400pt}}
\put(221,249){\makebox(0,0)[r]{\small$10^2$}}
\put(1010.0,249.0){\rule[-0.200pt]{4.818pt}{0.400pt}}
\put(241.0,275.0){\rule[-0.200pt]{2.409pt}{0.400pt}}
\put(1020.0,275.0){\rule[-0.200pt]{2.409pt}{0.400pt}}
\put(241.0,309.0){\rule[-0.200pt]{2.409pt}{0.400pt}}
\put(1020.0,309.0){\rule[-0.200pt]{2.409pt}{0.400pt}}
\put(241.0,327.0){\rule[-0.200pt]{2.409pt}{0.400pt}}
\put(1020.0,327.0){\rule[-0.200pt]{2.409pt}{0.400pt}}
\put(241.0,335.0){\rule[-0.200pt]{4.818pt}{0.400pt}}
\put(221,335){\makebox(0,0)[r]{\small $10^3$}}
\put(1010.0,335.0){\rule[-0.200pt]{4.818pt}{0.400pt}}
\put(241.0,361.0){\rule[-0.200pt]{2.409pt}{0.400pt}}
\put(1020.0,361.0){\rule[-0.200pt]{2.409pt}{0.400pt}}
\put(241.0,395.0){\rule[-0.200pt]{2.409pt}{0.400pt}}
\put(1020.0,395.0){\rule[-0.200pt]{2.409pt}{0.400pt}}
\put(241.0,413.0){\rule[-0.200pt]{2.409pt}{0.400pt}}
\put(1020.0,413.0){\rule[-0.200pt]{2.409pt}{0.400pt}}
\put(241.0,421.0){\rule[-0.200pt]{4.818pt}{0.400pt}}
\put(221,421){\makebox(0,0)[r]{\small $10^4$}}
\put(1010.0,421.0){\rule[-0.200pt]{4.818pt}{0.400pt}}
\put(241.0,447.0){\rule[-0.200pt]{2.409pt}{0.400pt}}
\put(1020.0,447.0){\rule[-0.200pt]{2.409pt}{0.400pt}}
\put(241.0,481.0){\rule[-0.200pt]{2.409pt}{0.400pt}}
\put(1020.0,481.0){\rule[-0.200pt]{2.409pt}{0.400pt}}
\put(241.0,499.0){\rule[-0.200pt]{2.409pt}{0.400pt}}
\put(1020.0,499.0){\rule[-0.200pt]{2.409pt}{0.400pt}}
\put(241.0,507.0){\rule[-0.200pt]{4.818pt}{0.400pt}}
\put(221,507){\makebox(0,0)[r]{\small $10^5$}}
\put(1010.0,507.0){\rule[-0.200pt]{4.818pt}{0.400pt}}
\put(241.0,533.0){\rule[-0.200pt]{2.409pt}{0.400pt}}
\put(1020.0,533.0){\rule[-0.200pt]{2.409pt}{0.400pt}}
\put(241.0,567.0){\rule[-0.200pt]{2.409pt}{0.400pt}}
\put(1020.0,567.0){\rule[-0.200pt]{2.409pt}{0.400pt}}
\put(241.0,585.0){\rule[-0.200pt]{2.409pt}{0.400pt}}
\put(1020.0,585.0){\rule[-0.200pt]{2.409pt}{0.400pt}}
\put(241.0,593.0){\rule[-0.200pt]{4.818pt}{0.400pt}}
\put(221,593){\makebox(0,0)[r]{\small $10^6$}}
\put(1010.0,593.0){\rule[-0.200pt]{4.818pt}{0.400pt}}
\put(241.0,619.0){\rule[-0.200pt]{2.409pt}{0.400pt}}
\put(1020.0,619.0){\rule[-0.200pt]{2.409pt}{0.400pt}}
\put(241.0,653.0){\rule[-0.200pt]{2.409pt}{0.400pt}}
\put(1020.0,653.0){\rule[-0.200pt]{2.409pt}{0.400pt}}
\put(241.0,671.0){\rule[-0.200pt]{2.409pt}{0.400pt}}
\put(1020.0,671.0){\rule[-0.200pt]{2.409pt}{0.400pt}}
\put(241.0,679.0){\rule[-0.200pt]{4.818pt}{0.400pt}}
\put(221,679){\makebox(0,0)[r]{\small $10^7$}}
\put(1010.0,679.0){\rule[-0.200pt]{4.818pt}{0.400pt}}
\put(241.0,163.0){\rule[-0.200pt]{0.400pt}{4.818pt}}
\put(241,122){\makebox(0,0){\small $10^{-8}$}}
\put(241.0,659.0){\rule[-0.200pt]{0.400pt}{4.818pt}}
\put(275.0,163.0){\rule[-0.200pt]{0.400pt}{2.409pt}}
\put(275.0,669.0){\rule[-0.200pt]{0.400pt}{2.409pt}}
\put(320.0,163.0){\rule[-0.200pt]{0.400pt}{2.409pt}}
\put(320.0,669.0){\rule[-0.200pt]{0.400pt}{2.409pt}}
\put(343.0,163.0){\rule[-0.200pt]{0.400pt}{2.409pt}}
\put(343.0,669.0){\rule[-0.200pt]{0.400pt}{2.409pt}}
\put(354.0,163.0){\rule[-0.200pt]{0.400pt}{4.818pt}}
\put(354,122){\makebox(0,0){\small $10^{-7}$}}
\put(354.0,659.0){\rule[-0.200pt]{0.400pt}{4.818pt}}
\put(388.0,163.0){\rule[-0.200pt]{0.400pt}{2.409pt}}
\put(388.0,669.0){\rule[-0.200pt]{0.400pt}{2.409pt}}
\put(432.0,163.0){\rule[-0.200pt]{0.400pt}{2.409pt}}
\put(432.0,669.0){\rule[-0.200pt]{0.400pt}{2.409pt}}
\put(456.0,163.0){\rule[-0.200pt]{0.400pt}{2.409pt}}
\put(456.0,669.0){\rule[-0.200pt]{0.400pt}{2.409pt}}
\put(466.0,163.0){\rule[-0.200pt]{0.400pt}{4.818pt}}
\put(466,122){\makebox(0,0){\small $10^{-6}$}}
\put(466.0,659.0){\rule[-0.200pt]{0.400pt}{4.818pt}}
\put(500.0,163.0){\rule[-0.200pt]{0.400pt}{2.409pt}}
\put(500.0,669.0){\rule[-0.200pt]{0.400pt}{2.409pt}}
\put(545.0,163.0){\rule[-0.200pt]{0.400pt}{2.409pt}}
\put(545.0,669.0){\rule[-0.200pt]{0.400pt}{2.409pt}}
\put(568.0,163.0){\rule[-0.200pt]{0.400pt}{2.409pt}}
\put(568.0,669.0){\rule[-0.200pt]{0.400pt}{2.409pt}}
\put(579.0,163.0){\rule[-0.200pt]{0.400pt}{4.818pt}}
\put(579,122){\makebox(0,0){\small $10^{-5}$}}
\put(579.0,659.0){\rule[-0.200pt]{0.400pt}{4.818pt}}
\put(613.0,163.0){\rule[-0.200pt]{0.400pt}{2.409pt}}
\put(613.0,669.0){\rule[-0.200pt]{0.400pt}{2.409pt}}
\put(658.0,163.0){\rule[-0.200pt]{0.400pt}{2.409pt}}
\put(658.0,669.0){\rule[-0.200pt]{0.400pt}{2.409pt}}
\put(681.0,163.0){\rule[-0.200pt]{0.400pt}{2.409pt}}
\put(681.0,669.0){\rule[-0.200pt]{0.400pt}{2.409pt}}
\put(692.0,163.0){\rule[-0.200pt]{0.400pt}{4.818pt}}
\put(692,122){\makebox(0,0){\small $10^{-4}$}}
\put(692.0,659.0){\rule[-0.200pt]{0.400pt}{4.818pt}}
\put(726.0,163.0){\rule[-0.200pt]{0.400pt}{2.409pt}}
\put(726.0,669.0){\rule[-0.200pt]{0.400pt}{2.409pt}}
\put(771.0,163.0){\rule[-0.200pt]{0.400pt}{2.409pt}}
\put(771.0,669.0){\rule[-0.200pt]{0.400pt}{2.409pt}}
\put(794.0,163.0){\rule[-0.200pt]{0.400pt}{2.409pt}}
\put(794.0,669.0){\rule[-0.200pt]{0.400pt}{2.409pt}}
\put(805.0,163.0){\rule[-0.200pt]{0.400pt}{4.818pt}}
\put(805,122){\makebox(0,0){\small $10^{-3}$}}
\put(805.0,659.0){\rule[-0.200pt]{0.400pt}{4.818pt}}
\put(839.0,163.0){\rule[-0.200pt]{0.400pt}{2.409pt}}
\put(839.0,669.0){\rule[-0.200pt]{0.400pt}{2.409pt}}
\put(883.0,163.0){\rule[-0.200pt]{0.400pt}{2.409pt}}
\put(883.0,669.0){\rule[-0.200pt]{0.400pt}{2.409pt}}
\put(906.0,163.0){\rule[-0.200pt]{0.400pt}{2.409pt}}
\put(906.0,669.0){\rule[-0.200pt]{0.400pt}{2.409pt}}
\put(917.0,163.0){\rule[-0.200pt]{0.400pt}{4.818pt}}
\put(917,122){\makebox(0,0){\small $10^{-2}$}}
\put(917.0,659.0){\rule[-0.200pt]{0.400pt}{4.818pt}}
\put(951.0,163.0){\rule[-0.200pt]{0.400pt}{2.409pt}}
\put(951.0,669.0){\rule[-0.200pt]{0.400pt}{2.409pt}}
\put(996.0,163.0){\rule[-0.200pt]{0.400pt}{2.409pt}}
\put(996.0,669.0){\rule[-0.200pt]{0.400pt}{2.409pt}}
\put(1019.0,163.0){\rule[-0.200pt]{0.400pt}{2.409pt}}
\put(1019.0,669.0){\rule[-0.200pt]{0.400pt}{2.409pt}}
\put(1030.0,163.0){\rule[-0.200pt]{0.400pt}{4.818pt}}
\put(1030,122){\makebox(0,0){\small $10^{-1}$}}
\put(1030.0,659.0){\rule[-0.200pt]{0.400pt}{4.818pt}}
\put(241.0,163.0){\rule[-0.200pt]{190.070pt}{0.400pt}}
\put(1030.0,163.0){\rule[-0.200pt]{0.400pt}{124.304pt}}
\put(241.0,679.0){\rule[-0.200pt]{190.070pt}{0.400pt}}
\put(41,421){\makebox(0,0){$\qquad\qquad n_{is}$}}
\put(635,61){\makebox(0,0){$\mu_{is}$}}
\put(295,290){\makebox(0,0)[l]{{\footnotesize No. of nucleotides}}}
\put(295,241){\makebox(0,0)[l]{{\footnotesize coding the CDRs}}}
\put(712,627){\makebox(0,0)[l]{{\footnotesize observed}}}
\put(712,593){\makebox(0,0)[l]{{\footnotesize mutation}}}
\put(712,559){\makebox(0,0)[l]{{\footnotesize rates}}}
\put(374,593){\makebox(0,0)[l]{model}}
\put(241.0,163.0){\rule[-0.200pt]{0.400pt}{124.304pt}}
\put(692,163){\line(0,1){516}}
\put(883,163){\line(0,1){516}}
\put(241,268){\line(1,0){789}}
\put(241,654){\usebox{\plotpoint}}
\multiput(241.00,652.93)(0.671,-0.482){9}{\rule{0.633pt}{0.116pt}}
\multiput(241.00,653.17)(6.685,-6.000){2}{\rule{0.317pt}{0.400pt}}
\multiput(249.00,646.93)(0.671,-0.482){9}{\rule{0.633pt}{0.116pt}}
\multiput(249.00,647.17)(6.685,-6.000){2}{\rule{0.317pt}{0.400pt}}
\multiput(257.00,640.93)(0.821,-0.477){7}{\rule{0.740pt}{0.115pt}}
\multiput(257.00,641.17)(6.464,-5.000){2}{\rule{0.370pt}{0.400pt}}
\multiput(265.00,635.93)(0.671,-0.482){9}{\rule{0.633pt}{0.116pt}}
\multiput(265.00,636.17)(6.685,-6.000){2}{\rule{0.317pt}{0.400pt}}
\multiput(273.00,629.93)(0.671,-0.482){9}{\rule{0.633pt}{0.116pt}}
\multiput(273.00,630.17)(6.685,-6.000){2}{\rule{0.317pt}{0.400pt}}
\multiput(281.00,623.93)(0.671,-0.482){9}{\rule{0.633pt}{0.116pt}}
\multiput(281.00,624.17)(6.685,-6.000){2}{\rule{0.317pt}{0.400pt}}
\multiput(289.00,617.93)(0.821,-0.477){7}{\rule{0.740pt}{0.115pt}}
\multiput(289.00,618.17)(6.464,-5.000){2}{\rule{0.370pt}{0.400pt}}
\multiput(297.00,612.93)(0.671,-0.482){9}{\rule{0.633pt}{0.116pt}}
\multiput(297.00,613.17)(6.685,-6.000){2}{\rule{0.317pt}{0.400pt}}
\multiput(305.00,606.93)(0.671,-0.482){9}{\rule{0.633pt}{0.116pt}}
\multiput(305.00,607.17)(6.685,-6.000){2}{\rule{0.317pt}{0.400pt}}
\multiput(313.00,600.93)(0.671,-0.482){9}{\rule{0.633pt}{0.116pt}}
\multiput(313.00,601.17)(6.685,-6.000){2}{\rule{0.317pt}{0.400pt}}
\multiput(321.00,594.93)(0.821,-0.477){7}{\rule{0.740pt}{0.115pt}}
\multiput(321.00,595.17)(6.464,-5.000){2}{\rule{0.370pt}{0.400pt}}
\multiput(329.00,589.93)(0.671,-0.482){9}{\rule{0.633pt}{0.116pt}}
\multiput(329.00,590.17)(6.685,-6.000){2}{\rule{0.317pt}{0.400pt}}
\multiput(337.00,583.93)(0.671,-0.482){9}{\rule{0.633pt}{0.116pt}}
\multiput(337.00,584.17)(6.685,-6.000){2}{\rule{0.317pt}{0.400pt}}
\multiput(345.00,577.93)(0.821,-0.477){7}{\rule{0.740pt}{0.115pt}}
\multiput(345.00,578.17)(6.464,-5.000){2}{\rule{0.370pt}{0.400pt}}
\multiput(353.00,572.93)(0.671,-0.482){9}{\rule{0.633pt}{0.116pt}}
\multiput(353.00,573.17)(6.685,-6.000){2}{\rule{0.317pt}{0.400pt}}
\multiput(361.00,566.93)(0.671,-0.482){9}{\rule{0.633pt}{0.116pt}}
\multiput(361.00,567.17)(6.685,-6.000){2}{\rule{0.317pt}{0.400pt}}
\multiput(369.00,560.93)(0.581,-0.482){9}{\rule{0.567pt}{0.116pt}}
\multiput(369.00,561.17)(5.824,-6.000){2}{\rule{0.283pt}{0.400pt}}
\multiput(376.00,554.93)(0.821,-0.477){7}{\rule{0.740pt}{0.115pt}}
\multiput(376.00,555.17)(6.464,-5.000){2}{\rule{0.370pt}{0.400pt}}
\multiput(384.00,549.93)(0.671,-0.482){9}{\rule{0.633pt}{0.116pt}}
\multiput(384.00,550.17)(6.685,-6.000){2}{\rule{0.317pt}{0.400pt}}
\multiput(392.00,543.93)(0.671,-0.482){9}{\rule{0.633pt}{0.116pt}}
\multiput(392.00,544.17)(6.685,-6.000){2}{\rule{0.317pt}{0.400pt}}
\multiput(400.00,537.93)(0.821,-0.477){7}{\rule{0.740pt}{0.115pt}}
\multiput(400.00,538.17)(6.464,-5.000){2}{\rule{0.370pt}{0.400pt}}
\multiput(408.00,532.93)(0.671,-0.482){9}{\rule{0.633pt}{0.116pt}}
\multiput(408.00,533.17)(6.685,-6.000){2}{\rule{0.317pt}{0.400pt}}
\multiput(416.00,526.93)(0.671,-0.482){9}{\rule{0.633pt}{0.116pt}}
\multiput(416.00,527.17)(6.685,-6.000){2}{\rule{0.317pt}{0.400pt}}
\multiput(424.00,520.93)(0.671,-0.482){9}{\rule{0.633pt}{0.116pt}}
\multiput(424.00,521.17)(6.685,-6.000){2}{\rule{0.317pt}{0.400pt}}
\multiput(432.00,514.93)(0.821,-0.477){7}{\rule{0.740pt}{0.115pt}}
\multiput(432.00,515.17)(6.464,-5.000){2}{\rule{0.370pt}{0.400pt}}
\multiput(440.00,509.93)(0.671,-0.482){9}{\rule{0.633pt}{0.116pt}}
\multiput(440.00,510.17)(6.685,-6.000){2}{\rule{0.317pt}{0.400pt}}
\multiput(448.00,503.93)(0.671,-0.482){9}{\rule{0.633pt}{0.116pt}}
\multiput(448.00,504.17)(6.685,-6.000){2}{\rule{0.317pt}{0.400pt}}
\multiput(456.00,497.93)(0.821,-0.477){7}{\rule{0.740pt}{0.115pt}}
\multiput(456.00,498.17)(6.464,-5.000){2}{\rule{0.370pt}{0.400pt}}
\multiput(464.00,492.93)(0.671,-0.482){9}{\rule{0.633pt}{0.116pt}}
\multiput(464.00,493.17)(6.685,-6.000){2}{\rule{0.317pt}{0.400pt}}
\multiput(472.00,486.93)(0.671,-0.482){9}{\rule{0.633pt}{0.116pt}}
\multiput(472.00,487.17)(6.685,-6.000){2}{\rule{0.317pt}{0.400pt}}
\multiput(480.00,480.93)(0.821,-0.477){7}{\rule{0.740pt}{0.115pt}}
\multiput(480.00,481.17)(6.464,-5.000){2}{\rule{0.370pt}{0.400pt}}
\multiput(488.00,475.93)(0.671,-0.482){9}{\rule{0.633pt}{0.116pt}}
\multiput(488.00,476.17)(6.685,-6.000){2}{\rule{0.317pt}{0.400pt}}
\multiput(496.00,469.93)(0.671,-0.482){9}{\rule{0.633pt}{0.116pt}}
\multiput(496.00,470.17)(6.685,-6.000){2}{\rule{0.317pt}{0.400pt}}
\multiput(504.00,463.93)(0.821,-0.477){7}{\rule{0.740pt}{0.115pt}}
\multiput(504.00,464.17)(6.464,-5.000){2}{\rule{0.370pt}{0.400pt}}
\multiput(512.00,458.93)(0.671,-0.482){9}{\rule{0.633pt}{0.116pt}}
\multiput(512.00,459.17)(6.685,-6.000){2}{\rule{0.317pt}{0.400pt}}
\multiput(520.00,452.93)(0.671,-0.482){9}{\rule{0.633pt}{0.116pt}}
\multiput(520.00,453.17)(6.685,-6.000){2}{\rule{0.317pt}{0.400pt}}
\multiput(528.00,446.93)(0.821,-0.477){7}{\rule{0.740pt}{0.115pt}}
\multiput(528.00,447.17)(6.464,-5.000){2}{\rule{0.370pt}{0.400pt}}
\multiput(536.00,441.93)(0.671,-0.482){9}{\rule{0.633pt}{0.116pt}}
\multiput(536.00,442.17)(6.685,-6.000){2}{\rule{0.317pt}{0.400pt}}
\multiput(544.00,435.93)(0.821,-0.477){7}{\rule{0.740pt}{0.115pt}}
\multiput(544.00,436.17)(6.464,-5.000){2}{\rule{0.370pt}{0.400pt}}
\multiput(552.00,430.93)(0.671,-0.482){9}{\rule{0.633pt}{0.116pt}}
\multiput(552.00,431.17)(6.685,-6.000){2}{\rule{0.317pt}{0.400pt}}
\multiput(560.00,424.93)(0.671,-0.482){9}{\rule{0.633pt}{0.116pt}}
\multiput(560.00,425.17)(6.685,-6.000){2}{\rule{0.317pt}{0.400pt}}
\multiput(568.00,418.93)(0.821,-0.477){7}{\rule{0.740pt}{0.115pt}}
\multiput(568.00,419.17)(6.464,-5.000){2}{\rule{0.370pt}{0.400pt}}
\multiput(576.00,413.93)(0.671,-0.482){9}{\rule{0.633pt}{0.116pt}}
\multiput(576.00,414.17)(6.685,-6.000){2}{\rule{0.317pt}{0.400pt}}
\multiput(584.00,407.93)(0.821,-0.477){7}{\rule{0.740pt}{0.115pt}}
\multiput(584.00,408.17)(6.464,-5.000){2}{\rule{0.370pt}{0.400pt}}
\multiput(592.00,402.93)(0.671,-0.482){9}{\rule{0.633pt}{0.116pt}}
\multiput(592.00,403.17)(6.685,-6.000){2}{\rule{0.317pt}{0.400pt}}
\multiput(600.00,396.93)(0.671,-0.482){9}{\rule{0.633pt}{0.116pt}}
\multiput(600.00,397.17)(6.685,-6.000){2}{\rule{0.317pt}{0.400pt}}
\multiput(608.00,390.93)(0.821,-0.477){7}{\rule{0.740pt}{0.115pt}}
\multiput(608.00,391.17)(6.464,-5.000){2}{\rule{0.370pt}{0.400pt}}
\multiput(616.00,385.93)(0.671,-0.482){9}{\rule{0.633pt}{0.116pt}}
\multiput(616.00,386.17)(6.685,-6.000){2}{\rule{0.317pt}{0.400pt}}
\multiput(624.00,379.93)(0.821,-0.477){7}{\rule{0.740pt}{0.115pt}}
\multiput(624.00,380.17)(6.464,-5.000){2}{\rule{0.370pt}{0.400pt}}
\multiput(632.00,374.93)(0.581,-0.482){9}{\rule{0.567pt}{0.116pt}}
\multiput(632.00,375.17)(5.824,-6.000){2}{\rule{0.283pt}{0.400pt}}
\multiput(639.00,368.93)(0.821,-0.477){7}{\rule{0.740pt}{0.115pt}}
\multiput(639.00,369.17)(6.464,-5.000){2}{\rule{0.370pt}{0.400pt}}
\multiput(647.00,363.93)(0.671,-0.482){9}{\rule{0.633pt}{0.116pt}}
\multiput(647.00,364.17)(6.685,-6.000){2}{\rule{0.317pt}{0.400pt}}
\multiput(655.00,357.93)(0.821,-0.477){7}{\rule{0.740pt}{0.115pt}}
\multiput(655.00,358.17)(6.464,-5.000){2}{\rule{0.370pt}{0.400pt}}
\multiput(663.00,352.93)(0.671,-0.482){9}{\rule{0.633pt}{0.116pt}}
\multiput(663.00,353.17)(6.685,-6.000){2}{\rule{0.317pt}{0.400pt}}
\multiput(671.00,346.93)(0.821,-0.477){7}{\rule{0.740pt}{0.115pt}}
\multiput(671.00,347.17)(6.464,-5.000){2}{\rule{0.370pt}{0.400pt}}
\multiput(679.00,341.93)(0.671,-0.482){9}{\rule{0.633pt}{0.116pt}}
\multiput(679.00,342.17)(6.685,-6.000){2}{\rule{0.317pt}{0.400pt}}
\multiput(687.00,335.93)(0.821,-0.477){7}{\rule{0.740pt}{0.115pt}}
\multiput(687.00,336.17)(6.464,-5.000){2}{\rule{0.370pt}{0.400pt}}
\multiput(695.00,330.93)(0.671,-0.482){9}{\rule{0.633pt}{0.116pt}}
\multiput(695.00,331.17)(6.685,-6.000){2}{\rule{0.317pt}{0.400pt}}
\multiput(703.00,324.93)(0.821,-0.477){7}{\rule{0.740pt}{0.115pt}}
\multiput(703.00,325.17)(6.464,-5.000){2}{\rule{0.370pt}{0.400pt}}
\multiput(711.00,319.93)(0.671,-0.482){9}{\rule{0.633pt}{0.116pt}}
\multiput(711.00,320.17)(6.685,-6.000){2}{\rule{0.317pt}{0.400pt}}
\multiput(719.00,313.93)(0.821,-0.477){7}{\rule{0.740pt}{0.115pt}}
\multiput(719.00,314.17)(6.464,-5.000){2}{\rule{0.370pt}{0.400pt}}
\multiput(727.00,308.93)(0.671,-0.482){9}{\rule{0.633pt}{0.116pt}}
\multiput(727.00,309.17)(6.685,-6.000){2}{\rule{0.317pt}{0.400pt}}
\multiput(735.00,302.93)(0.821,-0.477){7}{\rule{0.740pt}{0.115pt}}
\multiput(735.00,303.17)(6.464,-5.000){2}{\rule{0.370pt}{0.400pt}}
\multiput(743.00,297.93)(0.671,-0.482){9}{\rule{0.633pt}{0.116pt}}
\multiput(743.00,298.17)(6.685,-6.000){2}{\rule{0.317pt}{0.400pt}}
\multiput(751.00,291.93)(0.821,-0.477){7}{\rule{0.740pt}{0.115pt}}
\multiput(751.00,292.17)(6.464,-5.000){2}{\rule{0.370pt}{0.400pt}}
\multiput(759.00,286.93)(0.821,-0.477){7}{\rule{0.740pt}{0.115pt}}
\multiput(759.00,287.17)(6.464,-5.000){2}{\rule{0.370pt}{0.400pt}}
\multiput(767.00,281.93)(0.671,-0.482){9}{\rule{0.633pt}{0.116pt}}
\multiput(767.00,282.17)(6.685,-6.000){2}{\rule{0.317pt}{0.400pt}}
\multiput(775.00,275.93)(0.821,-0.477){7}{\rule{0.740pt}{0.115pt}}
\multiput(775.00,276.17)(6.464,-5.000){2}{\rule{0.370pt}{0.400pt}}
\multiput(783.00,270.93)(0.671,-0.482){9}{\rule{0.633pt}{0.116pt}}
\multiput(783.00,271.17)(6.685,-6.000){2}{\rule{0.317pt}{0.400pt}}
\multiput(791.00,264.93)(0.821,-0.477){7}{\rule{0.740pt}{0.115pt}}
\multiput(791.00,265.17)(6.464,-5.000){2}{\rule{0.370pt}{0.400pt}}
\multiput(799.00,259.93)(0.821,-0.477){7}{\rule{0.740pt}{0.115pt}}
\multiput(799.00,260.17)(6.464,-5.000){2}{\rule{0.370pt}{0.400pt}}
\multiput(807.00,254.93)(0.671,-0.482){9}{\rule{0.633pt}{0.116pt}}
\multiput(807.00,255.17)(6.685,-6.000){2}{\rule{0.317pt}{0.400pt}}
\multiput(815.00,248.93)(0.821,-0.477){7}{\rule{0.740pt}{0.115pt}}
\multiput(815.00,249.17)(6.464,-5.000){2}{\rule{0.370pt}{0.400pt}}
\multiput(823.00,243.93)(0.821,-0.477){7}{\rule{0.740pt}{0.115pt}}
\multiput(823.00,244.17)(6.464,-5.000){2}{\rule{0.370pt}{0.400pt}}
\multiput(831.00,238.93)(0.821,-0.477){7}{\rule{0.740pt}{0.115pt}}
\multiput(831.00,239.17)(6.464,-5.000){2}{\rule{0.370pt}{0.400pt}}
\multiput(839.00,233.93)(0.671,-0.482){9}{\rule{0.633pt}{0.116pt}}
\multiput(839.00,234.17)(6.685,-6.000){2}{\rule{0.317pt}{0.400pt}}
\multiput(847.00,227.93)(0.821,-0.477){7}{\rule{0.740pt}{0.115pt}}
\multiput(847.00,228.17)(6.464,-5.000){2}{\rule{0.370pt}{0.400pt}}
\multiput(855.00,222.93)(0.821,-0.477){7}{\rule{0.740pt}{0.115pt}}
\multiput(855.00,223.17)(6.464,-5.000){2}{\rule{0.370pt}{0.400pt}}
\multiput(863.00,217.93)(0.821,-0.477){7}{\rule{0.740pt}{0.115pt}}
\multiput(863.00,218.17)(6.464,-5.000){2}{\rule{0.370pt}{0.400pt}}
\multiput(871.00,212.93)(0.671,-0.482){9}{\rule{0.633pt}{0.116pt}}
\multiput(871.00,213.17)(6.685,-6.000){2}{\rule{0.317pt}{0.400pt}}
\multiput(879.00,206.93)(0.821,-0.477){7}{\rule{0.740pt}{0.115pt}}
\multiput(879.00,207.17)(6.464,-5.000){2}{\rule{0.370pt}{0.400pt}}
\multiput(887.00,201.93)(0.821,-0.477){7}{\rule{0.740pt}{0.115pt}}
\multiput(887.00,202.17)(6.464,-5.000){2}{\rule{0.370pt}{0.400pt}}
\multiput(895.00,196.93)(0.710,-0.477){7}{\rule{0.660pt}{0.115pt}}
\multiput(895.00,197.17)(5.630,-5.000){2}{\rule{0.330pt}{0.400pt}}
\multiput(902.00,191.93)(0.821,-0.477){7}{\rule{0.740pt}{0.115pt}}
\multiput(902.00,192.17)(6.464,-5.000){2}{\rule{0.370pt}{0.400pt}}
\multiput(910.00,186.93)(0.821,-0.477){7}{\rule{0.740pt}{0.115pt}}
\multiput(910.00,187.17)(6.464,-5.000){2}{\rule{0.370pt}{0.400pt}}
\multiput(918.00,181.93)(0.821,-0.477){7}{\rule{0.740pt}{0.115pt}}
\multiput(918.00,182.17)(6.464,-5.000){2}{\rule{0.370pt}{0.400pt}}
\multiput(926.00,176.93)(0.821,-0.477){7}{\rule{0.740pt}{0.115pt}}
\multiput(926.00,177.17)(6.464,-5.000){2}{\rule{0.370pt}{0.400pt}}
\multiput(934.00,171.93)(0.821,-0.477){7}{\rule{0.740pt}{0.115pt}}
\multiput(934.00,172.17)(6.464,-5.000){2}{\rule{0.370pt}{0.400pt}}
\multiput(942.00,166.93)(0.821,-0.477){7}{\rule{0.740pt}{0.115pt}}
\multiput(942.00,167.17)(6.464,-5.000){2}{\rule{0.370pt}{0.400pt}}
\end{picture}
\caption{\label{pic2}Predicted mutation rates vs.\ receptor coding lengths  
for an optimal immune response, in comparison to observed rates of somatic 
hypermutation and observed CDR coding lengths. 
}
\end{center}
\end{figure}
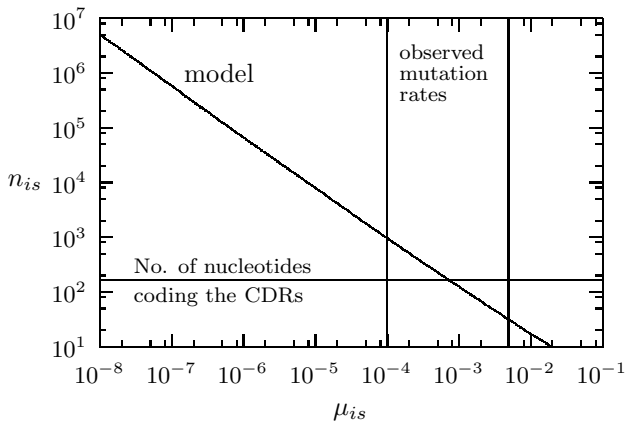
As Fig.\ \ref{pic2} shows, the model prediction agrees well with the observed 
somatic hypermutation rates and CDR receptor lengths.  

To summarize, the dynamics of the co-evolution of two coupled quasispecies 
has been studied. In particular this model was formulated to provide a simple
toy model for the co-adaptive system of viral evolution and immune adaptation.  
The model characterizes the different regimes of (co-)existence of viral and 
immune quasispecies and predicts the correct range of somatic mutation rates
in accordance with observation. Possible extensions of this work are numerous, 
as this is only a first account of basic principles of co-evolving quasispecies. 
Analytical approaches beyond the simple approximation presented here, as well as 
numerical extensions may provide a more accurate picture of the dynamics and 
further possibilities to relate to biological data. Further applications include
modeling HIV dynamics, e.g. by adding an overall decay rate representing the 
HIV-induced loss of $CD4^+$ T-cells.
\medskip

C.K. would like to thank the Stiftung der Deutschen Wirtschaft for financial 
support.


\begin{thebibliography}{10}
\bibitem[*]{email}
Email address: bornholdt@theo-physik.uni-kiel.de
\bibitem{eigen71}
M.~Eigen, Naturwissenschaften {\bf 58}, 465 (1971).
\bibitem{eigen79}
M.\ Eigen and P.~Schuster, {\em The Hypercycle - A Principle of Natural 
Self-Organization} (Springer Verlag, Berlin, 1979). 
\bibitem{nowak89}
M.~Nowak and P.~Schuster, J.\ Theoret.\ Biol.\ {\bf 137}, 375 (1989).
\bibitem{taraz92}
P.~Tarazona, Phys. Rev. A {\bf 45}, 6038 (1992).
\bibitem{bonh93}
S.~Bonhoeffer and P.~Stadler, J. Theoret. Biol. {\bf 164}, 359 (1993).
\bibitem{baake97}
E.\ Baake, M.\ Baake, and H.\ Wagner, Phys. Rev. Lett. {\bf 78}, 559 (1997).
\bibitem{altm01}
S.~Altmeyer and J.~McCaskill, Phys. Rev. Lett. {\bf 86}, 5819 (2001).
\bibitem{nilsson00}
M.~Nilsson and N.~Snoad, Phys. Rev. Lett. {\bf 84}, 191 (2000).
\bibitem{wilke99}
C.\ Wilke, C. Ronnewinkel, and T. Martinetz, Phys. Rep. {\bf 349}, 395 (2001).
\bibitem{harris99}
R.~S. Harris, Q.~Kong, and N.~Maizels, Mutation Research {\bf 436}, 157 (1999).
\bibitem{meyer01}
M.\ Meyer-Herrman, A.\ Deutsch, and M.\ Or-Guil, 
J.\ Theor.\ Biol.\ {\bf 210}, 265 (2001). 
\bibitem{ganeshan97}
S.~Ganeshan, R.~E. Dickover, B.~Korber, Y.~Bryson, and S.~Wolinsky,
J. Virol. {\bf 71}, 663 (1997).
\bibitem{allen00}
T.~Allen et~al., Nature {\bf 407}, 386 (2000).
\bibitem{jones76}
B.~Jones, R.~Enns, and S.~Rangnekar, Bull. Math. Biol. {\bf 38}, 15 (1976).
\bibitem{thompson74}
C.~Thompson and J.~McBride, Math. Biosci. {\bf 21}, 127 (1974).
\bibitem{roitt}
I.~Roitt, {\em Essential Immunolgy} (Blackwell Scientific Publ., 1994).
\bibitem{aigner}
A.~Aigner and S.~Neumann, {\em Immunchemie} (Fischer, 1997). 
\bibitem{weigert70}
M.~Weigert, I.~Cesari, S.~J. Yonkovich, and M.~Cohn, 
Nature {\bf 228}, 1045 (1970).
\bibitem{nossal92}
G.~Nossal, Cell {\bf 68}, 1 (1992).
\bibitem{mckean84}
D.~McKean et~al., Proc. Natl. Acad. Sci. USA {\bf 81}, 3180 (1984).
\bibitem{berek88}
C.~Berek, C.~Milstein, Immunol. Rev. {\bf 105}, 5 (1988).
\bibitem{levy89}
N.~S. Levy, U.~V. Malipiero, S.~G. Lebecque, and P.~J. Gearhart,
J. Exp. Med. {\bf 169}, 2007 (1989).
\bibitem{doerner97}
T.~D\"orner et~al., J. Immunol. {\bf 158}, 2779 (1997).
\bibitem{diaz98}
M.~F. Diaz, M.~Flajnik, Immunol. Rev. {\bf 162}, 13 (1998).
\bibitem{kasahara98}
M.~Kasahara, Immunol. Rev. {\bf 166}, 159 (1998).
\bibitem{wilson92}
M.~Wilson et~al., EMBO J. {\bf 11}, 4337 (1992).
\bibitem{reynaud95}
C.-A. Reynaud, C.~Garcia, W.~Hein, and J.-C. Weill, 
Cell {\bf 80}, 115 (1995).
\bibitem{green98}
N.~S. Green, M.~Lin, and M.~D. Scharff, 
Immunol. Rev. {\bf 162}, 77 (1998).
\end{thebibliography}
\end{document}